\documentclass[aps,prd,10pt,nofootinbib,twocolumn,superscriptaddress,floatfix,notitlepage]{revtex4-1}
\pdfoutput=1

\usepackage{graphicx}
\usepackage{amsmath,amsfonts,amssymb}
\usepackage{color}
\usepackage[breaklinks,colorlinks,urlcolor=blue,citecolor=blue,linkcolor=magenta]{hyperref}
\usepackage{verbatim}
\usepackage{enumitem}
\usepackage{aas_macros}

\newcommand{\Msun}{M_\odot}
\newcommand{\td}{{\rm d}}

\newcommand{\be}{\begin{equation}}
\newcommand{\ee}{\end{equation}}
\newcommand{\bea}{\begin{equation} \begin{aligned}}
\newcommand{\eea}{\end{aligned} \end{equation}}

\def\lsim{\mathrel{\raise.3ex\hbox{$<$\kern-.75em\lower1ex\hbox{$\sim$}}}}
\def\gsim{\mathrel{\raise.3ex\hbox{$>$\kern-.75em\lower1ex\hbox{$\sim$}}}}

\begin{document}

\title{Eccentricity effects on the SMBH background}

\author{Juhan Raidal}
\email{juhan.raidal@kbfi.ee}
\affiliation{Keemilise ja Bioloogilise F\"u\"usika Instituut, R\"avala pst. 10, 10143 Tallinn, Estonia}

\author{Juan Urrutia}
\email{juan.urrutia@kbfi.ee}
\affiliation{Keemilise ja Bioloogilise F\"u\"usika Instituut, R\"avala pst. 10, 10143 Tallinn, Estonia}
\affiliation{Departament of Cybernetics, Tallinn University of Technology, Akadeemia tee 21, 12618 Tallinn, Estonia}

\author{Ville Vaskonen}
\email{ville.vaskonen@pd.infn.it}
\affiliation{Keemilise ja Bioloogilise F\"u\"usika Instituut, R\"avala pst. 10, 10143 Tallinn, Estonia}
\affiliation{Dipartimento di Fisica e Astronomia, Universit\`a degli Studi di Padova, Via Marzolo 8, 35131 Padova, Italy}
\affiliation{Istituto Nazionale di Fisica Nucleare, Sezione di Padova, Via Marzolo 8, 35131 Padova, Italy}

\author{Hardi Veerm\"ae}
\email{hardi.veermae@cern.ch}
\affiliation{Keemilise ja Bioloogilise F\"u\"usika Instituut, R\"avala pst. 10, 10143 Tallinn, Estonia}

\begin{abstract}
We studied how eccentricity affects the gravitational wave (GW) spectrum from supermassive black hole (SMBH) binaries. We developed a fast and accurate semi-analytic method for computing the GW spectra, the distribution for the spectral fluctuations and the correlations between different frequencies. As GW emission circularizes binaries, the suppression of the signal strength due to eccentricity is relevant for signals from wider binaries emitting at lower frequencies. Such a feature is present in the signal observed at pulsar timing arrays. We found that when orbital decay of the SMBH binaries is driven by GWs only, the shape of the observed signal preferred highly eccentric binaries $\langle e \rangle_{2\,{\rm nHz}} = 0.83^{+0.04}_{-0.05}$. However, when environmental effects were included, the initial eccentricity could be significantly lowered, yet the scenario with purely circular binaries was still mildly disfavored.
\end{abstract}

\maketitle

\section{Introduction}

The pulsar timing array (PTA) collaborations have recently found compelling evidence for the Hellings-Down quadrupolar correlation~\citep{Hellings:1983fr} within the previously observed common-spectrum stochastic process~\citep{NANOGrav:2023gor, EPTA:2023fyk, EPTA:2023sfo, Reardon:2023gzh, Zic:2023gta, Reardon:2023zen, Xu:2023wog}. This indicates that the observed signal likely originates from a stochastic background of gravitational waves (GWs), the measurement of which marks a significant milestone in GW astronomy. Such a GW background at nHz frequencies can be naturally sourced by a population of inspiraling supermassive black hole (SMBH) binaries~\citep{NANOGrav:2023hde,EPTA:2023xxk,Ellis:2023dgf,Ellis:2023oxs}. 

The naive expectation of the GW background from circular SMBH binaries, a power-law with a spectral index of $13/3$~\citep{Phinney:2001di}, does not provide a good fit to the data, and less steep spectra are preferred. The fit improves significantly once the stochastic nature of the background and energy loss via environmental effects are included~\citep{Ellis:2023dgf}. The additional energy loss also aids in solving the final parsec problem~\citep{Begelman:1980vb}, as the energy loss shortens the timescale of tightly bound binary formation to reasonable (sub-Hubble time) lengths~\citep{Ellis:2024wdh}.

The SMBH binaries are formed in mergers of galaxies~\citep{Begelman:1980vb} and typically have very high initial eccentricities~\citep{Fastidio:2024crh}. Whether the binary retains its high initial eccentricity when it reaches parsec separations, where the emission of GWs becomes relevant, is unclear and has been studied with simulations and semi-analytic models~\citep{Dotti:2006ef,2017MNRAS.466.1170M, DOrazio:2021kob, Berczik:2006tz,2015ApJ...810...49V,Gualandris:2022kxh,Kelley:2017lek,Fastidio:2024crh}.

If the binary is in a gas-rich environment and forms a circumbinary disk, the eccentricity of the binary evolves by interactions with the gas. For circular circumbinary disks, it is expected that, at the pericenter, the SMBH orbits faster than the disk and the disk tends to slow it down. In contrast, at the apocenter, the SMBH orbits slower than the disk and the disk tends to speed it up. This circularizes the binary making the SMBH coalesce with a negligible eccentricity~\citep{Dotti:2006ef}. However, more recent simulations suggest that angular momentum transfer between the binary and the accretion disk could maintain large eccentricities and increase the separation~\citep{2017MNRAS.466.1170M}. A transition between the circular and the precessing disk has been found to happen at around $e \sim 0.4$~\citep{DOrazio:2021kob,2021ApJ...909L..13Z}, and all initial eccentricities $e \gsim \mathcal{O}(0.1)$ end up at $e \sim 0.4$ while for $e \lsim \mathcal{O}(0.1)$ the circumbinary disk remains circular and circularizes the orbits.

On the other hand, if the merger happens in a gas-poor environment, loss-cone scattering can drive the binary to sub-parsec separation in a sub-Hubble time if the loss-cone is refilled sufficiently fast~\citep{Berczik:2006tz,2015ApJ...810...49V}. The loss-cone scattering phase is stochastic and its overall effect on the eccentricity distribution depends on the number of stellar encounters~\citep{Nasim:2020kpw, Rawlings:2023dcr}, on the initial eccentricity~\citep{Gualandris:2022kxh}, on the geometry of the stellar background~\citep{Rawlings:2023dcr} and the mass ratio of the SMBH binary~\citep{Gualandris:2022kxh}. Overall, these effects can circularize the orbit or, for comparable mass binaries, preserve the high eccentricity. During hardening, eccentricity might increase in minor mergers but is reduced in major ones, permitting large eccentricities of up to $e\sim 0.8$~\citep{Gualandris:2022kxh} at the onset of the GW-driven phase. Furthermore, the results of the Illustris simulation indicate that SMBH binaries can preserve a very high eccentricity up to the point they enter the PTA band~\citep{Kelley:2017lek, Fastidio:2024crh}.

Although a clear consensus is lacking about the eccentricity distribution of SMBH binaries entering the PTA frequency band, it is possible that their eccentricities can be relatively high. Thus, in light of the recent PTA data, it is not only important to understand how the eccentricity would affect the nHz GW spectrum from SMBHs, but a good theoretical picture of the latter will assist us in resolving the properties of the SMBH binary population.

In this paper, we studied how eccentricities of SMBH binaries affect the nHz GW background. Large eccentricities in the GW-dominated stage of binary evolution lead to a suppression of the GW signal, due to faster orbital decay~\citep{Enoki:2006kj, Sesana:2013wja, Huerta:2015pva, Chen:2016zyo, Bi:2023tib}. Since GW-emission circularizes the orbits, this effect is more relevant at lower frequencies where larger average eccentricities are expected. We showed that the eccentricity-induced suppression improves the fit to the NANOGrav 15-year (NG15) data~\citep{NANOGrav:2023gor}. Furthermore, we found that the effect of eccentricity on the GW background was remarkably similar to that expected from environmental effects~\citep{Ellis:2023dgf}.

\section{GW emission of eccentric SMBHs}

Unlike circular binaries, eccentric binaries emit GWs at multiple frequencies. A binary with a given orbital frequency, $f_{\rm b}$, emits GWs at the rest frame frequencies 
\be
    f_{n,{\rm r}} =(1+z)f_n = n f_{\rm b} \,, 
\ee
where $n\geq1$ is the harmonic number and $f_n$ is the measured GW frequency of the $n$th harmonic. For circular binaries, only the $n=2$ harmonic is present. An eccentric SMBH binary emits GWs in the $n$th harmonic with power~\citep{Peters:1963ux}\footnote{Geometric units $c = G = 1$ are used throughout the paper.}
\bea\label{eq:En}
    \frac{\td E_n}{\td t_r}&=\frac{32}{5}(2\pi\mathcal{M}f_{\rm b})^{10/3}g_n(e)\\
    &= \frac{\td E_c}{\td t_r}g_n(e),
\eea
where $\mathcal{M}$ denotes the chirp mass and $\td E_c/\td t_r$ is the power emitted by a circular binary with the same binary parameters $\mathcal{M}$, $f_{\rm b}$ and $z$. The observed time $t$ is related to the time $t_r$ in the rest frame of the binary by $\td t = (1+z) \td t_r$. The relative power radiated in the $n$th harmonic is
\bea\label{eq:gn}
    &g_n(e) = \frac{n^4}{32} \Bigg[ \frac{4}{3n^2}J_n^2(ne) + \bigg(J_{n-2}(ne)-2eJ_{n-1}(ne) \\
    &\hspace{8pt} + \frac{2}{n}J_n(ne)+2eJ_{n+1}(ne)-J_{n+2}(ne)\bigg)^{\!2} \\
    &\hspace{8pt} + (1-e^2)\bigg(J_{n-2}(ne)-2J_n(ne)+J_{n+2}(ne)\bigg)^{\!2} \,\Bigg] ,
\eea
where $J_n$ is the $n$th Bessel function of the first kind. For circular binaries, only the second harmonic contributes $g_n(0) = \delta_{2n}$. The total power emitted by a binary across all frequencies is
\be\label{eq:Etot}
     \frac{\td E_{\rm GW}}{\td t_r} = \frac{\td E_c}{\td t_r}\mathcal{F}(e) \,,
\ee
where
\be
     \mathcal{F}(e) = \sum_{n=1}^\infty g_n(e)=\frac{1+73/24e^2+37/96e^4}{(1-e^2)^{7/2}} \,.
\ee
GW emission by an eccentric binary, $e>0$, is enhanced compared to emission by a circular binary by $\mathcal{F}(e) > 1$.

The increase in the total emitted GW power does not necessarily lead to an increase in the GW background because the increase in the emitted power accelerates the orbital decay of the binary. The orbital dynamics of an eccentric binary are described by the evolution of its orbital frequency and eccentricity~\citep{Peters:1963ux}
\bea \label{eq:orbital}
    &\frac{\td \ln f_{\rm b}}{\td t_r} = \frac{96}{5} \mathcal{M}^{5/3} (2\pi f_{\rm b})^{8/3} \mathcal{F}(e) \,, \\
    &\frac{\td e}{\td t_r} = -\frac{1}{15} \mathcal{M}^{5/3} (2\pi f_{\rm b})^{8/3} \mathcal{G}(e) \,,
\eea
where
\be \label{eq:Ge}
    \mathcal{G}(e) = \frac{304e+121e^3}{(1-e^2)^{5/2}} \,.
\ee
From this, the observed residence time of a binary at orbital frequency $f_{\rm b}$ is
\bea\label{eq:residence1}
    \frac{\td t}{\td \ln f_{\rm b}} 
    &= \frac{5}{96} \frac{1+z}{\mathcal{M}^{5/3}} (2\pi f_{\rm b})^{-8/3} \frac{1}{\mathcal{F}(e)}\\
    &= \frac{\td t_c}{\td \ln f_{\rm b}} \frac{1}{\mathcal{F}(e)} \,,
\eea
where $\td t_c/\td \ln f_{\rm b}$ is the observed residence time of the corresponding circular binary (i.e., at $e=0$) and the factor $1+z$ arises from the conversion of the binary rest frame time $t_r$ to the time $t$ in the observer frame.

\section{Eccentricity distribution}

The orbital evolution of the SMBH binaries is very slow compared to the PTA observation times, so the binaries can be treated as having a constant orbital frequency and eccentricity from an observational point of view. On the other hand, Eq.~\eqref{eq:orbital} gives
\be \label{eq:dedlnf}
    \frac{\td e}{\td \ln{f_{\rm b}}} = -\frac{1}{288} \frac{\mathcal{G}(e)}{\mathcal{F}(e)} \,,
\ee
indicating a strong relation between the eccentricity of a binary and its frequency. Binaries with higher orbital frequencies are more likely to have lower eccentricities because the GW emission circularizes their orbits. By~\eqref{eq:dedlnf}, eccentricity evolution in a GW-driven regime can be described by
\be\label{eq:dedlnf2}
    f_{\rm b} \propto \left(1-e^2\right)^{\frac{3}{2}} e^{-\frac{18}{19}} \left(304 + 121e^2\right)^{-\frac{1305}{2299}} .
\ee
Importantly, it is universal, that is, it does not depend on binary masses, if there are no additional environmental effects.

\begin{figure}
    \centering    
    \includegraphics[width=0.85\columnwidth]{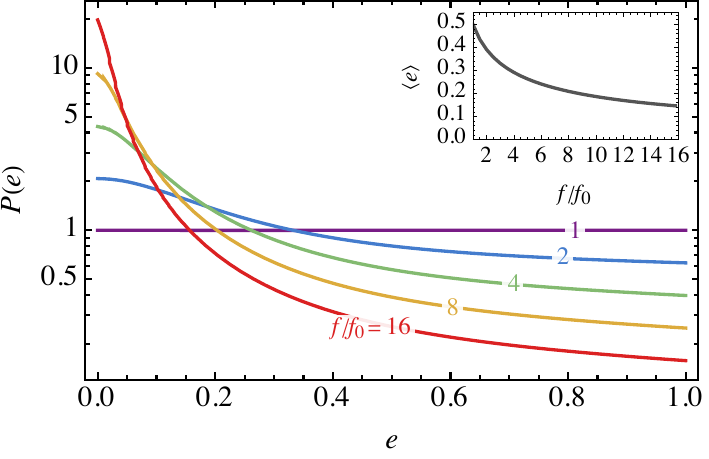}
    \vspace{-2mm}
    \caption{Evolution of an eccentricity distribution set to be uniform at $f=f_0$. The inset shows the evolution of the expectation value of $e$.}
    \label{fig:Pe}
\end{figure}

We started with fixing the initial eccentricity distribution, $P(e,f_0) = P_0(e)$ at some initial orbital frequency $f_0$ (see also ~\citet{Huerta:2015pva} and \citet{Chen:2016zyo}). Eq.~\eqref{eq:dedlnf2} gives the evolution of eccentricity with frequency, $e = \tilde{e}(e_0,f_{\rm b}/f_0)$. This can be inverted to give $e_0 = \tilde{e}(e,f_0/f_{\rm b})$ and, since probability is conserved, the distribution of eccentricities at frequency $f$ is
\be \label{eq:Pe}
    P(e|f_{\rm b}) = \frac{\td \tilde{e}(e,f_0/f_{\rm b})}{\td e}P_0(\tilde{e}(e,f_0/f_{\rm b})) \,.
\ee
For the initial distribution, we consider a power law,
\be
    P_0(e) = (\gamma+1) e^{\gamma} \,,
\ee
with $\gamma>-1$. Although it is a generic phenomenological ansatz, it has some physical motivation as, for instance, $\gamma=1$ corresponds to the "thermal" distribution~\citep{1919MNRAS..79..408J}. In the following, we characterized the distribution using the mean initial eccentricity, which is related to the parameter $\gamma$ as 
\be
    \langle e \rangle_{f_0} = \frac{1+\gamma}{2+\gamma} \,,
\ee
or inversely,  $\gamma = - (1-2 \langle e \rangle_{f_0})/(1-\langle e \rangle_{f_0})$. An example of how the evolution of eccentricity affects its distribution at different frequencies is shown in Fig.~\ref{fig:Pe} for a uniform initial distribution ($\gamma=0$).

\section{Mean GW spectrum}

The mean GW spectrum generated by eccentric SMBH binaries is
\bea \label{eq:mean1}
    \langle\Omega(f)\rangle = \frac{1}{\rho_c} \int \td \lambda\,& \td e\, \td f_{\rm b}\, \sum_{n=1}^\infty \delta\!\left(f_{\rm b} - \frac{f_{\rm r}}{n}\right) \\ 
    &\times  P(e|f_{\rm b})  \frac{1+z}{4\pi D_L^2} \frac{\td E_n}{\td \ln{f_{\rm b}}} \,,
\eea
where $\rho_c$ is the critical energy density, $D_L$ is the source luminosity distance and the delta function picks binaries whose $n$th harmonic contributes to the observed frequency $f = f_{\rm r}/(1+z)$. The differential merger rate of BHs in the observer reference frame is
\be \label{eq:diffmergerrate}
    \td \lambda = \td \mathcal{M} \td \eta \td z \frac{1}{1+z} \frac{\td V_c}{\td z} \frac{\td R_{\rm BH}}{\td \mathcal{M} \td \eta}\,,
\ee
where $\eta$ is the symmetric mass ratio, $V_c$ the comoving volume, and $R_{\rm BH}$ the comoving BH merger rate density. 

Following~\citet{Ellis:2023dgf}, we used the halo merger rate $R_h$ arising from the extended Press-Schechter formalism~\citep{Press:1973iz,Bond:1990iw,1993MNRAS.262..627L},\footnote{We computed the variance of smoothed mass fluctuations following~\citet{Eisenstein:1997ik} and used the values of the cosmological parameters inferred from the CMB observations~\citep{Planck:2018vyg}.} and converted it to the BH merger rate through a halo mass-BH mass relation
\bea \label{eq:mergerrate}
    \frac{\td R_{\rm BH}}{\td m_1 \td m_2} = \,p_{\rm BH} \int & \td M_{{\rm v},1} \td M_{{\rm v},2} \, \frac{\td R_h}{\td M_{{\rm v},1} \td M_{{\rm v},2}}\\
    & \times \prod_{j=1,2} \frac{\td P(m_j|M_*(M_{{\rm v},j},z))}{\td m_j}  \,, 
\eea
where $m_{j}$ are the masses of the merging BHs, $M_{{\rm v},j}$ are the virial masses of their host halos and $p_{\rm BH} \le 1$ combines the SMBH occupation fraction in galaxies with the efficiency for the BHs to merge following the merging of their host halos. Due to the GW background being generated by binaries with a narrow range of SMBH masses and redshifts, $p_{\rm BH}$ was treated as a constant free parameter. We estimated the mass relation using the observed halo mass-stellar mass relation $M_*(M_{{\rm v},j},z)$ from~\citet{Girelli:2020goz} and the global fit stellar mass-BH mass relation from~\citet{Ellis:2024wdh}. The latter is given by
\be
    \frac{\td P(m|M_*)}{\td \log_{10} \!m} = \mathcal{N}\bigg(\!\log_{10} \!\frac{m}{M_\odot} \bigg| a + b \log_{10} \!\frac{M_*}{10^{11}M_\odot},\sigma\bigg) \, ,
\ee
with $a = 8.6$, $b = 0.8$ and $\sigma = 0.8$, where $\mathcal{N}(x|\bar x,\sigma)$ denotes the probability density of a Gaussian distribution with mean $\bar x$ and variance $\sigma^2$. We assume that eccentricity does not significantly affect the merger rate.

The residence time at each emitting frequency is reduced for eccentric binaries, leading to an energy emitted per logarithmic frequency interval of
\bea \label{eq:residence2}
    \frac{\td E_n}{\td \ln{f_{\rm b}}} = \frac{\td t_c}{\td \ln f_{\rm b}} \frac{1}{1+z} \frac{\td E_c}{\td t_r} \frac{g_n(e)}{\mathcal{F}(e)} \,.
\eea
Now, using~\eqref{eq:residence2}, the mean GW abundance~\eqref{eq:mean1} can be expressed as
\bea \label{eq:mean2}
    \langle\Omega(f)\rangle 
    &\!= \!\!\int \!\! \td \lambda \td e \sum_{n=1}^\infty P(e|f_{\rm b}) \frac{g_n(e)}{\mathcal{F}(e)} \Omega_c^{(1)}\! \frac{\td t_c}{\td \ln{f_{\rm b}}} \bigg|_{f_{\rm b} = \frac{f_r}{n}} \\
    &\!= \!\!\int \td \lambda \, \mathcal{E}(f_{\rm r}) \, \Omega_c^{(1)} \frac{\td t_c}{\td \ln f_{\rm b}}\bigg|_{f_{\rm b} = \frac{f_{\rm r}}{2}},
\eea
where we defined
\be \label{eq:Omegac1}
    \Omega_c^{(1)} \equiv \frac{1}{4\pi D_L^2\rho_c}\frac{\td E_c}{\td t_r}
\ee
denoting the contribution from a single circular binary, and\footnote{The factor of $(2/n)^{2/3}$ arises from changing orbital frequencies from $f_{\rm b}=f_{\rm r}/n$ to $f_{\rm b}=f_{\rm r}/2$ in Eq.~\eqref{eq:mean2}.}
\be \label{eq:mathE}
    \mathcal{E}(f_{\rm r}) \equiv \sum_{n=1}^\infty \left[\frac{2}{n}\right]^{2/3} \int \td e  \,P(e|f_{\rm r}/n) \frac{g_n(e)}{\mathcal{F}(e)} \,.
\ee
which quantifies how non-zero eccentricities modify the mean GW spectrum of circular binaries. For circular binaries, $P(e) = \delta(e)$ and $\mathcal{E}(f_{\rm r}) = 1$. Eq.~\eqref{eq:mathE} agrees with~\citet[Eq.~(36)]{Huerta:2015pva} for delta distributions $P(e) = \delta(e - e_0)$. We remarked that the second line of~\eqref{eq:mean2} holds only for inspiraling binaries when the binary evolution is driven purely by GW emission and environmental effects can be neglected.

\begin{figure}
    \centering    
    \includegraphics[width=0.85\columnwidth]{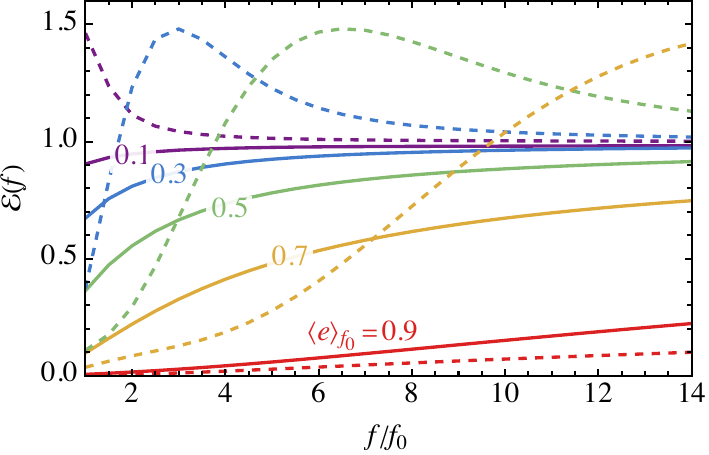}
    \caption{Function $\mathcal{E}$ defined in \eqref{eq:mathE} that describes the modification of the mean GW spectrum caused by the binary eccentricities shown for power-law (solid) and delta function (dashed) initial eccentricity distributions and different values of mean eccentricity of the initial distribution fixed at $f=f_0$.}
    \label{fig:Enf}
\end{figure}

From the initial eccentricity distribution and its evolution, we evaluated \eqref{eq:mathE} which shows how eccentricity affects the mean GW spectrum. In Fig.~\ref{fig:Enf}, we showed $\mathcal{E}(f)$ for different initial eccentricity distributions. Since GW emission circularizes orbits as the orbital frequency increases, $\mathcal{E}(f)$ reaches unity at $f\gg f_0$. High initial eccentricities suppress the spectrum at low frequencies. For a power-law $P_0(e)$ we find that $\mathcal{E}(f)$ increases monotonously with $f$ whereas for $P_0(e) = \delta(e-e_0)$ (or for narrow peaked distributions), it increases until it gets mildly larger and starts to approach unity from above.

\section{Statistics of GW spectra}
\label{sec:gwspectrum}

Given that the binary separation and the eccentricity evolve slowly when compared to an orbital period, the observed GW spectrum of a {\it single} binary can be expressed as a sum of individual harmonics:
\be
    \Omega^{(1)}(f) = \Omega^{(1)}_c \sum_{n\geq 1} g_n(e) f \,\delta\left(f - \frac{n}{2}f_2\right)\,.
\ee
Thus, the statistical properties of the GW spectrum of SMBH binaries can be inferred from the distribution of $\Omega_c$ and $e$ (or an equivalent set of parameters) at a given $f_2$ (or observed orbital frequency $f_1$) and the number of SMBH binaries in a given range of $f_2$. In the following, we considered the distribution
\bea \label{eq:P1circ}
    P^{(1)}(\Omega_c,e|f_2) = &\left(\frac{\td N}{\td \ln f_2}\right)^{-1} \int  \td \lambda P(e|f_{\rm b})  \\
    &\times \frac{\td t_c}{\td \ln f_{\rm b}} \delta(\Omega_c - \Omega^{(1)}_{c}) \bigg|_{f_{\rm b} = \frac{f_{2,r}}{2}} \,,
\eea
where the number of SMBHs per a logarithmic $f_2$ interval is
\be
    \frac{\td N}{\td \ln f_2} = \int \td \lambda \frac{\td t_c}{\td \ln f_{\rm b}}\bigg|_{f_{\rm b} = \frac{f_{2,r}}{2}}\,.
\ee
The mean~\eqref{eq:mean2} is recovered as
\bea
    \langle\Omega(f)\rangle = \int \td e \,& \td \Omega_c \sum_n \frac{\Omega_c g_n(e)}{\mathcal{F}(e)} \\
    &\times \frac{\td N}{\td \ln f_2} P^{(1)}(\Omega_c,e|f_2) \big|_{f_2 = \frac{2f}{n}} \,.
\eea
Higher order correlation functions can be computed similarly from $P^{(1)}(\Omega_c,e|f_2)$.

\subsection{Distribution at a frequency}

The GW spectrum in a given frequency bin $f_j$ can be computed by integrating over eccentricities $e$. The distribution of $\Omega_c$ and $e$ at a given $f_2$ can then be converted into a distribution of contributions $\Omega$ from individual binaries at a given frequency $f$,
\bea \label{eq:P1semi-analytic}
    P^{(1)}(\Omega|f) 
    =& \left( \frac{\td N}{\td \ln f} \right)^{-1} \int \td e \,\td\lambda \sum_{n=1}^\infty \frac{P(e|f_{\rm b})}{\mathcal{F}(e)} \\
    &\times \frac{\td t_c}{\td\ln{f_{\rm b}}} \delta(\Omega - \Omega_c^{(1)} g_n(e)) \bigg|_{f_{\rm b} = \frac{f_r}{n}} \,,
\eea
where 
\be
    \frac{\td N}{\td \ln f} = \int \td \lambda \frac{\td t_c}{\td \ln f_{\rm b}}\bigg|_{f_{\rm b} = \frac{f_{r}}{2}} \,.
\ee
The only difference with the approach developed in~\citet{Ellis:2023owy} and \citet{Ellis:2023dgf} is that now the expression includes a sum over the harmonics $n$, an integral over eccentricity $e$, with the appropriate $g_n(e)$ and $\mathcal{F}(e)$ factors, and the distribution of eccentricities. Consequently, the derivation of the statistical properties of the GW spectrum presented below has a sizeable overlap with these references.

We computed the distribution of the total GW abundance in frequency bins, $P(\Omega|f_j)$, by dividing the sources into strong sources, above some threshold $\Omega > \Omega_{\rm thr}$, and weak sources, $\Omega < \Omega_{\rm thr}$. The strong sources were modelled via Monte-Carlo (MC) sampling~\eqref{eq:P1circ} in each frequency bin. The expected number of binaries in a frequency interval $(f_j,f_{j+1})$ is
\be
    N_j = \int_{f_j}^{f_{j+1}} \td \ln f  \frac{\td N}{\td \ln f} \,,
\ee 
and the number of strong sources is $N_{{\rm S},j} = p_{{\rm S},j} N_j \ll N_j$, where
\be
    p_{{\rm S},j} = \int_{\Omega>\Omega_{\rm thr}} \!\!\! \td \Omega \,P^{(1)}(\Omega|f_j) \,.
\ee
We sampled $10^5$ realisations of the strong source contribution and chose $\Omega_{\rm thr}$ so that the number of strong sources is $N_{{\rm S},j} = 50$. This number of strong sources was sufficient to accurately model the high-amplitude fluctuations in the spectrum (for more details, see \citet{Ellis:2023dgf}). The distribution of $\Omega$ from individual sources has scales approximately as $P^{(1)}(\Omega|f)\propto \Omega^{-5/2}$ at large $\Omega$~\citep{Ellis:2023owy}. This long tail is inherited by $P(\Omega)$~\citep{Ellis:2023dgf}
\be\label{eq:tail}
    P(\Omega) \!\!\stackrel{\Omega \gg \langle\Omega\rangle}{\approx}\!\!P^{(1)}\!\left[(\Omega - \langle\Omega\rangle)\ln\frac{f_{j+1}}{f_j} \right]  N_j  \ln\!\frac{f_{j+1}}{f_j},
\ee
where $\langle\Omega\rangle$ is the mean of $\Omega$ in a given frequency bin.

\begin{figure}
    \centering    
    \includegraphics[width=0.85\columnwidth]{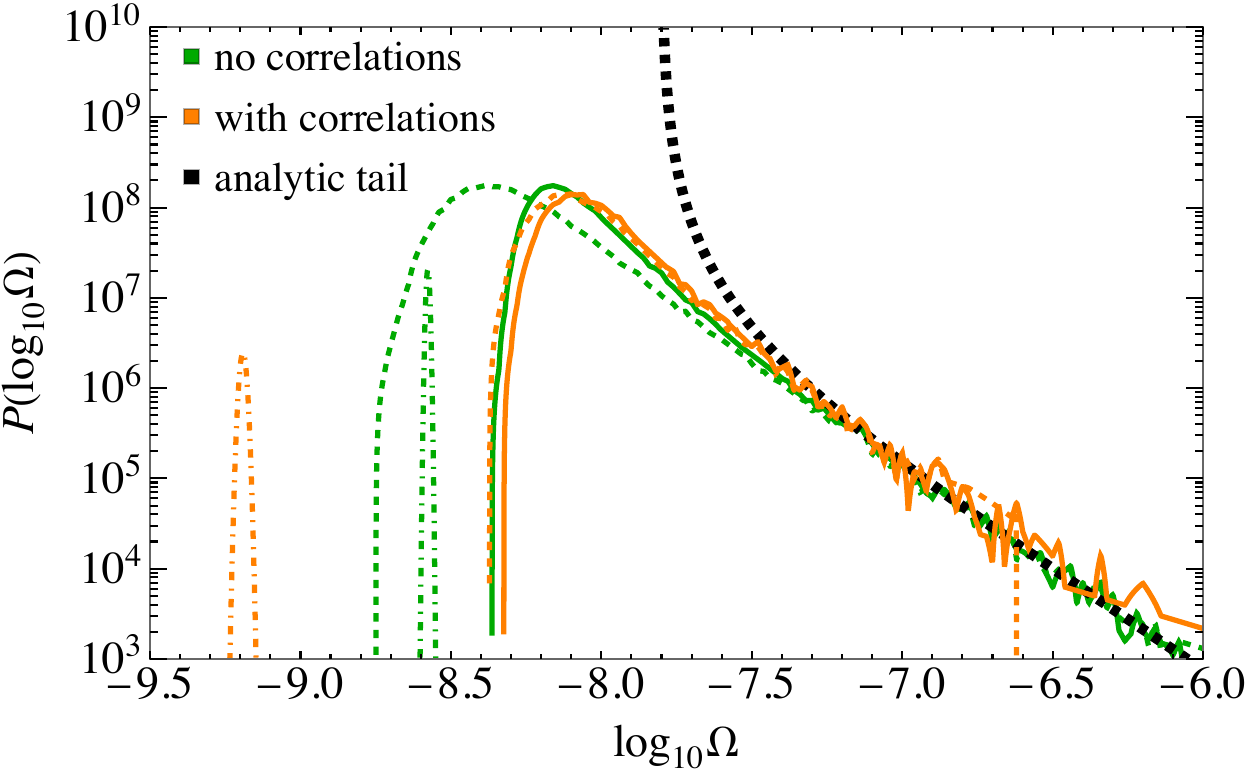}
    \vspace{-2mm}
    \caption{Distribution of the total GW abundance in the seventh NG15 frequency bin for the best-fit eccentricity model found in Sec.~\ref{sec:results}. The green (orange) curve shows the distribution obtained using the no-correlation (with correlations) method with $10^5$ ($10^4$) realisations. The thick black dashed curve shows the analytical estimate of the long tail of the distribution of individual binaries. The dashed and dot-dashed curves show the contributions from strong and weak sources respectively.}
    \label{fig:tail}
\end{figure}

\begin{figure}
    \centering    
    \includegraphics[width=0.85\columnwidth]{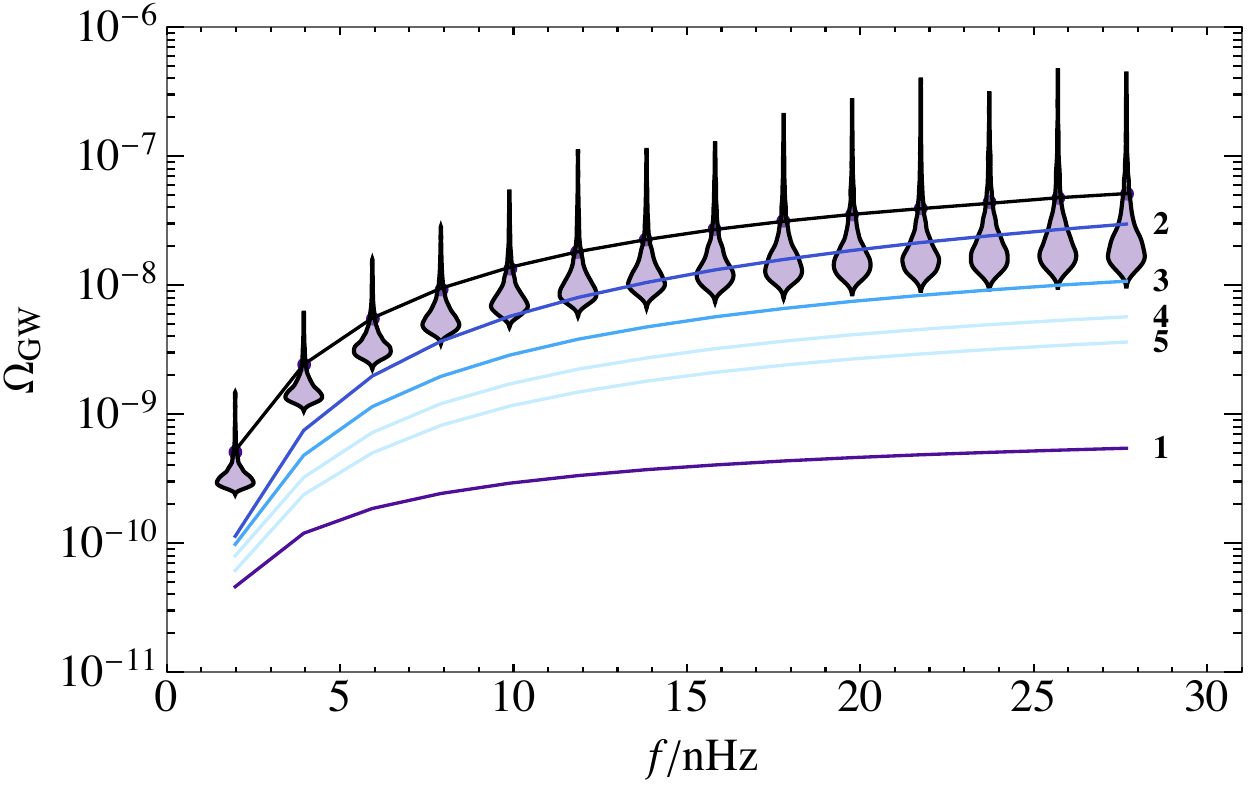}
    \vspace{-2mm}
    \caption{Best-fit eccentricity spectrum found in Sec.~\ref{sec:results} (violins) and the means of the first 5 harmonics (solid curves) contributing towards the mean of the spectrum (solid black curve).}
    \label{fig:meanconts}
\end{figure}

The weak sources did not need to be treated with an MC approach, as their distribution obeys the central limit theorem and does not inherit the long $\Omega$ tail of $ P^{(1)}(\Omega|f)$ due to the cutoff at the division between strong and weak sources. Thus, the GW signal from the weak sources follows Gaussian distribution. Moreover, as this distribution tends to be narrow, we estimated their effect on the total distribution by simply shifting the contribution from the strong sources by the mean of the weak source distribution,
\be
    \langle \Omega_{\rm W}(f_j) \rangle = \int_{\Omega<\Omega_{\rm thr}} \!\!\! \,\td \Omega \,P^{(1)}(\Omega|f_j) \Omega \,.
\ee

An example of the distribution $P(\log_{10}\Omega|f_j)$ of $\Omega$ in the frequency bin $f_j\in [13.8 - 15.8]\, {\rm nHz}$ is shown in Fig.~\ref{fig:tail} along with the individual components contributing to it. The dash-dotted lines depict the contribution of weak sources and the dashed lines show the contribution from strong sources. The green and orange lines correspond to the distributions obtained, respectively, by omitting and including correlations between different bins. In the latter case, individual spectra of the 50 strongest binaries characterised by $\Omega_c$, $e$ in a small interval of $f_2$ were drawn from the distribution $P^{(1)}(\Omega_c,e|f_2)$ given in~\eqref{eq:P1circ} (see sec.~\ref{sec:correlations}), while the uncorrelated case generated sets of 50 realizations of $\Omega$ from $P^{(1)}(\Omega|f)$ in a given frequency bin $f_i$. Due to this difference in binning, the number of strong sources contributing to a given frequency was smaller in the former case, which lead to a stronger signal from the remaining weak sources -- the green dashed-dotted distribution in Fig.~\ref{fig:tail} is seen to peak at a larger 
value of $\Omega$ than the orange dashed-dotted distribution. Nevertheless, as expected, both approaches gave the same distribution of $\Omega$ even when both the weak and strong sources were combined. The analytic tail \eqref{eq:tail}, shown by the dashed line, was reproduced well with the MC results.

The best-fit eccentric spectrum (found in Sec.~\ref{sec:results}) and the first five harmonics contributing toward its mean can be seen in Fig.~\ref{fig:meanconts}. Note that, due to the large initial mean eccentricity, there is still significant mean eccentricity present in the system even in the last bin. This leads to relevant contributions to the mean from multiple harmonics, not just the second harmonic. The mean, shown by the black curve, is noticeably above the median of the violins because of the long high-$\Omega$ tail of the distributions seen in Fig.~\ref{fig:tail}. 

\begin{figure*}
    \centering    
    \includegraphics[width=0.9\textwidth]{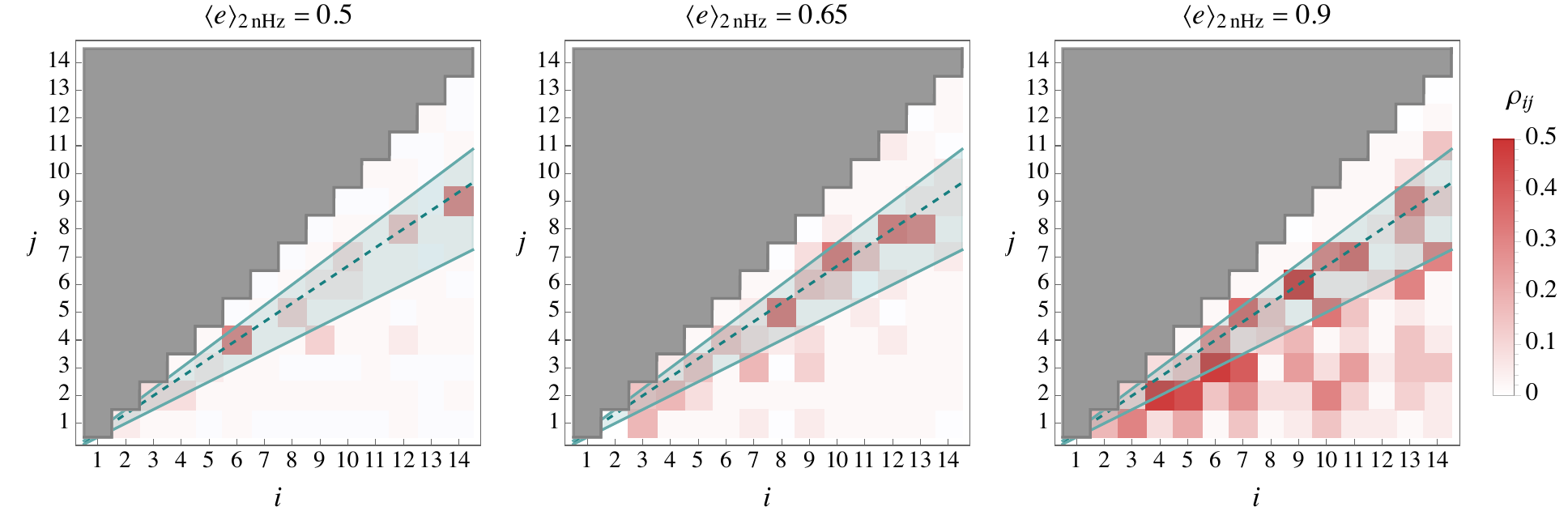}
    \caption{Correlations between frequency bins for spectra generated with $10^4$ realisations, $p_{\rm BH}=1$ and differing initial mean eccentricities at orbital frequency $f_{\rm b}={2\,{\rm nHz}}$. The correlations get stronger as eccentricity increases, with the strongest correlations found for frequency ratios $2/3$ (along the blue dashed line) and most of the correlated bins are expected to lie between the ratios $3/4$ and $1/2$ (shaded blue area).}
    \label{fig:correlations}
\end{figure*}

\subsection{Correlations between frequencies}
\label{sec:correlations}

Since eccentric binaries emit GWs at several frequencies, they can contribute to more than one frequency bin and consequently make them correlated. To study these correlations, we needed to sample both $\Omega_c$ and $e$ from~\eqref{eq:P1circ}. We, again, computed the total GW abundance in frequency bins, $P(\Omega|f_j)$, by dividing the sources into strong and weak sources. Here the threshold was on $\Omega_c$ instead of $\Omega$ and, since we needed to fix $f_2$ instead of the frequencies of the data bins, we binned $f_2$ so that each output (NG15) frequency bin was divided into $10$ smaller frequency bins and both higher and lower frequencies were added to account for signal "leaking" into the PTA sensitivity range. We chose $f_2$ from each of the small input bins and sampled simultaneously $\Omega_c$ and $e$ from~\eqref{eq:P1circ}. The expected number of binaries in a frequency interval $(f_{2,j},f_{2,j+1})$ is
\be
    N_j = \int_{f_j}^{f_{j+1}} \td \ln f_2 \frac{\td N}{\td \ln f_2} \,,
\ee 
and the number of strong sources is $N_{{\rm S},j} = p_{{\rm S},j} N_j \ll N_j$, where
\be
    p_{{\rm S},j} = \int_{\Omega_c^{(1)}>\Omega_{\rm thr}} \!\!\! \td e \,\td \Omega_c \,P^{(1)}(\Omega_c,e|f_j) \,.
\ee
The emission by each strong binary was binned into the data bins to get the total contribution from the strong sources:
\be
    \Omega_{\rm S}(f_j) = \frac{1}{\ln{(f_{j+1}/f_j)}} \sum_{k=1}^{N_{\rm S}} \sum_{n\in N_j} \frac{\Omega_{c,k} g_n(e_k)}{\mathcal{F}(e_k)} \,,
\ee
where $k$ labels the strong binaries and $n\in N_j=\{2f_j/f_{2,k},\dots,2f_{j+1}/f_{2,k}\}$ are the harmonics of each strong binary that contribute to the frequency bin $f_j$.

The correlation between the frequency bins $i$ and $j$ is given by
\be
    \rho_{ij}=\frac{\langle\Omega(f_i)\Omega(f_j)\rangle- \langle\Omega(f_i)\rangle \langle\Omega(f_j)\rangle}{\sigma_i\sigma_j}\, ,
\ee
where $\sigma_{i}=\sqrt{\langle\Omega(f_i)^2\rangle-\langle{\Omega}(f_i)\rangle^2}$. As shown in Fig.~\ref{fig:tail}, the contribution from weak sources (dash-dotted yellow curve) is negligible in this approach. This is because for the computation of the correlations we considered 10 times more bins of the second harmonic frequency $f_2$ than in the computation based on the observed frequency, $f$, but we kept the number of strong sources in each bin the same. Consequently, the strong sources almost fully dominate the background and we can neglect the weak sources.\footnote{Correlations between the frequency bins of the weak source background can be described by a multivariate Gaussian
\bea
    P&(\Omega_{\rm W}(f_1),\Omega_{\rm W}(f_2),\dots,\Omega_{\rm W}(f_j)) \\
    &\approx \frac{\exp\left[-(\boldsymbol{\Omega}_{\rm W}-\langle\boldsymbol{\Omega}_{\rm W}\rangle)^T\boldsymbol{\Sigma}^{-1}(\boldsymbol{\Omega}_{\rm W}- \langle\boldsymbol{\Omega}_{\rm W})\rangle\right]}{\sqrt{(2\pi)^j|\boldsymbol{\Sigma}|}}\,,
\eea
where $j$ is the number of bins, $\boldsymbol{\Sigma}$ is the covariance matrix and $\langle\boldsymbol{\Omega}_{\rm W}\rangle$ is the vector of mean $\Omega_{\rm W}$ in each bin. The covariance matrix of the multivariate normal distribution is then $\Sigma_{ij} \equiv \sigma_i\sigma_j\rho_{ij}$ where $\sigma_{i}=\sqrt{\langle\Omega_{{\rm W}}(f_i)^2\rangle-\langle{\Omega}_{{\rm W}}(f_i)\rangle^2}$.} 

Fig.~\ref{fig:correlations} shows correlations between $\Omega(f_i)$ in different frequency bins for various initial eccentricities in a SMBH model with $p_{\rm BH} = 1$. The correlations depend on the amount of eccentricity present in the binary population, with the strongest correlations following a line of frequency ratio $2/3$. This ratio dominates because the spectrum of moderately eccentric binaries tends to be dominated by the 2nd and the 3rd harmonics. When the 4th harmonic is included, then most of the correlations are expected to be found in the region between frequency ratios $3/4$ and $1/2$, highlighted by the turquoise band in Fig.~\ref{fig:correlations}. We found that the correlations can be negligible already for relatively large initial eccentricities $\langle e\rangle_{2 \rm nHz} \lsim 0.5$, while notable correlations required very high initial eccentricities $\langle e\rangle_{2 \rm nHz} = 0.9$. In such cases, the strongly correlated bins seen can appear away from the $1/2 - 3/4$ ratio band due to the non-vanishing contribution from higher harmonics.

\section{Environmental effects}

In addition to GW emission, SMBH binaries may lose energy through dissipative environmental effects. The change in the binaries binding energy $E$ is then
\be
    \Dot{E} = -\Dot{E}_{\rm GW}-\Dot{E}_{\text{env}} \,.
\ee
The impact of environmental effects is not fully known. Though it is generally expected that encounters with stars and drag due to circumbinary gas disks shorten binary inspiral times ~\citep{Tang:2017eiz,Armitage:2002uu,Merritt:2013awa}, some also argue that environmental effects can expand binary orbits~\citep{Munoz:2018tnj}. Following~\citet{Ellis:2023dgf}, we introduced environmental effects via a phenomenological power-law ansatz.

The characteristic timescales for GW emission and environmental effect energy loss are $t_{\rm GW}=|E|/\Dot{E}_{\rm GW}=4\tau$ and $t_{\text{env}}=|E|/\Dot{E}_{\text{env}}$, where $\tau$ is the coalescence time of the binary assuming only GW emission. Following Eq.~\eqref{eq:Etot}, we see that
\be
    t_{\rm GW}=\frac{|E|}{\Dot{E}_{\rm GW}}=\frac{|E|}{\Dot{E}^c_{\rm GW}\mathcal{F}(e)}=\frac{4\tau_c}{\mathcal{F}(e)},
\ee
where $\tau_c = (5/256) (1+z) (\pi f_r)^{-8/3} \mathcal{M}^{-5/3}$ is the observed coalescence time of a circular binary, $\tau_c=\lim_{e\to 0}\tau$. By computing the time derivative of the binding energy $E = -\left(\pi f_r \right)^{2/3} \mathcal{M}^{5/3}/2$, the impact of environmental effects to the residence time of binaries at each logarithmic frequency can then be found to be
\bea
     \frac{\td t}{\td \ln f_r} &= \frac23 \left(t_{\rm GW}^{-1} + t_{\rm env}^{-1}\right)^{-1} \\
     &= \frac23 \frac{t^c_{\rm GW}}{\mathcal{F}(e)} \left[1 + \frac{t^c_{\rm GW}}{t_{\rm env}}\frac{1}{\mathcal{F}(e)}\right]^{-1} ,
\eea
giving
\be\label{eq:ecceffects}
    \frac{\td E_n}{\td \ln f_r} = \frac{\td E_n}{\td \ln f_r}\bigg|_{t_{\text{env}}\to \infty} \left[1 + \frac{t^c_{\rm GW}}{t_{\rm env}}\frac{1}{\mathcal{F}(e)}\right]^{-1} \,.
\ee
To leading order, the timescale of the environmental effects can be approximated with a power-law treatment
\be
    \frac{t_{\rm env}}{t_{\rm GW}^c}=\left(\frac{f_r}{f_{\rm GW}}\right)^\alpha \, ,
\ee
where $f_{\rm GW}$ is the reference frequency above which GW emission becomes the dominant form of energy loss. We parametrize $f_{\rm GW}$ as
\be
    f_{\rm GW}(\mathcal{M},\eta,z)=f_{\rm ref}\left(\frac{\mathcal{M}}{10^9 \Msun}\right)^{-\beta} .
\ee

The modified orbital evolution due to the introduction of environmental effects also has an effect on the evolution of eccentricity distributions as a function of orbital frequency, with Eq.~\eqref{eq:dedlnf} now becoming
\be\label{eq:eccenvevolve}
    \frac{\td e}{\td \ln f_{\rm b}} = -\frac{1}{288}\frac{\mathcal{G}(e)}{\mathcal{F}(e)}\left[1 + \frac{t^c_{\rm GW}}{t_{\rm env}}\frac{1}{\mathcal{F}(e)}\right]^{-1} .
\ee
Here we implicitly assumed that the environmental effects do not change the evolution of binary eccentricity, i.e., $\td e/\td t_r$ is still given by Eq.~\eqref{eq:orbital}.\footnote{A similar approach is adopted in~\citet[Eq.~(31)]{Chen:2016zyo}, where eccentricity evolution below a threshold frequency was turned off. Instead of a sharp cutoff, Eq.~\eqref{eq:eccenvevolve} adopts a more gradual suppression of the eccentricity evolution.}

The evolution of the eccentricity distribution now depends on the masses of the binary components. Our numerical estimates suggested that environmental effects affected the eccentricity evolution of light-to-moderate binaries. As an explicit example, we considered a scenario with the environmental parameters fixed to the best-fit values obtained in~\citet{Ellis:2024wdh} and an initial eccentricity of $\langle e \rangle_{2\,{\rm nHz}}=0.80$. Independently of the binary mass, the mean eccentricity in the last bin was $\langle e \rangle_{28\,{\rm nHz}}=0.40$ when no environmental effects were present. When environmental effects were included $\langle e \rangle_{28\,{\rm nHz}}=0.41$ for binaries with $\mathcal{M}=10^{10}\Msun$, $\langle e \rangle_{28\,{\rm nHz}}=0.67$ for binaries with $\mathcal{M}=10^{8}\Msun$ and the mean eccentricity of light binaries, $\mathcal{M}\lsim 10^{6}\Msun$, remained practically constant at all bins indicating that orbital evolution is almost entirely dominated by environmental effects. It follows that, since light binaries exhibit no evolution of mean eccentricity as a function of orbital frequency, their contribution to the final spectrum is uniformly suppressed and becomes even less significant.

In both Eq.~\eqref{eq:ecceffects} and \eqref{eq:eccenvevolve}, the environmental effect terms are scaled by a $\mathcal{F}^{-1}(e)$ factor. Since $\mathcal{F}(e)\geq 1$, environmental effects are suppressed for highly-eccentric binaries and, for most of the parameter space, the evolution of binaries is either eccentricity or environment-dominated.

\section{Fit to the PTA data}
\label{sec:results}

\begin{figure}
    \centering    
    \includegraphics[width=0.7\columnwidth]{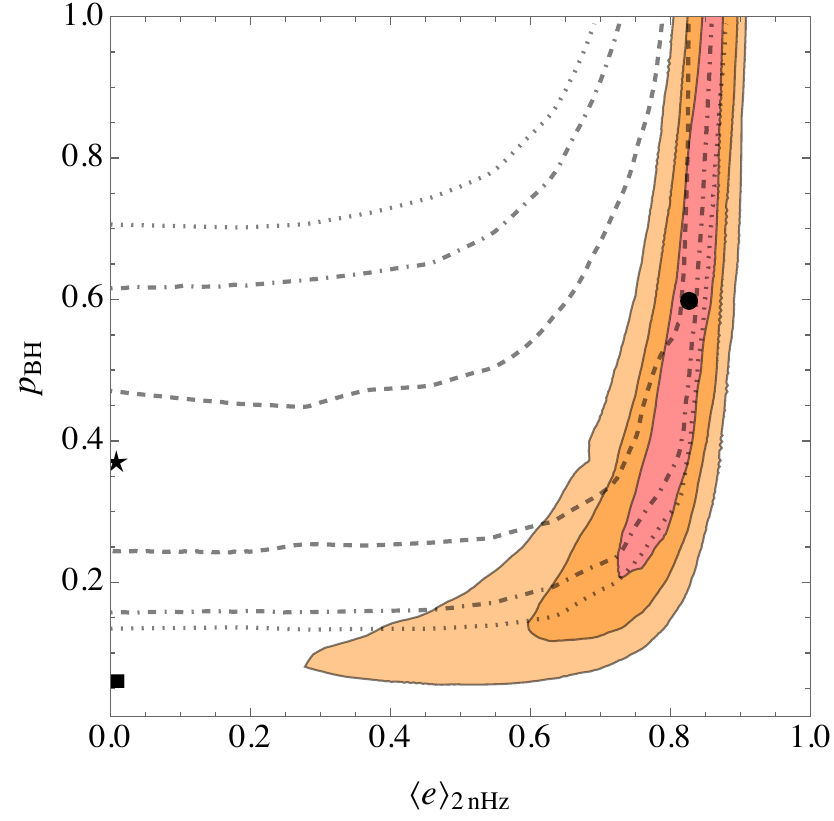}
    \caption{Fit of the eccentric SMBH binary model (solid) and a model with fixed environmental parameters ($f_{\rm ref}=30 \rm nHz, \alpha=8/3,\beta=5/8$) with added eccentricity (dashed). Note that the confidence levels are shown for each fit separately, with the ratio between the $1\sigma$ contours of the purely eccentric and environmental+eccentricity model being $\approx 5$. The black circle denotes the best-fit point in the model including eccentricities, while the black square/star indicate the best-fit points for circular binaries with environmental effects omitted/included.}
    \label{fig:likelihood}
\end{figure}

\begin{figure}
    \centering    
    \includegraphics[width=0.85\columnwidth]{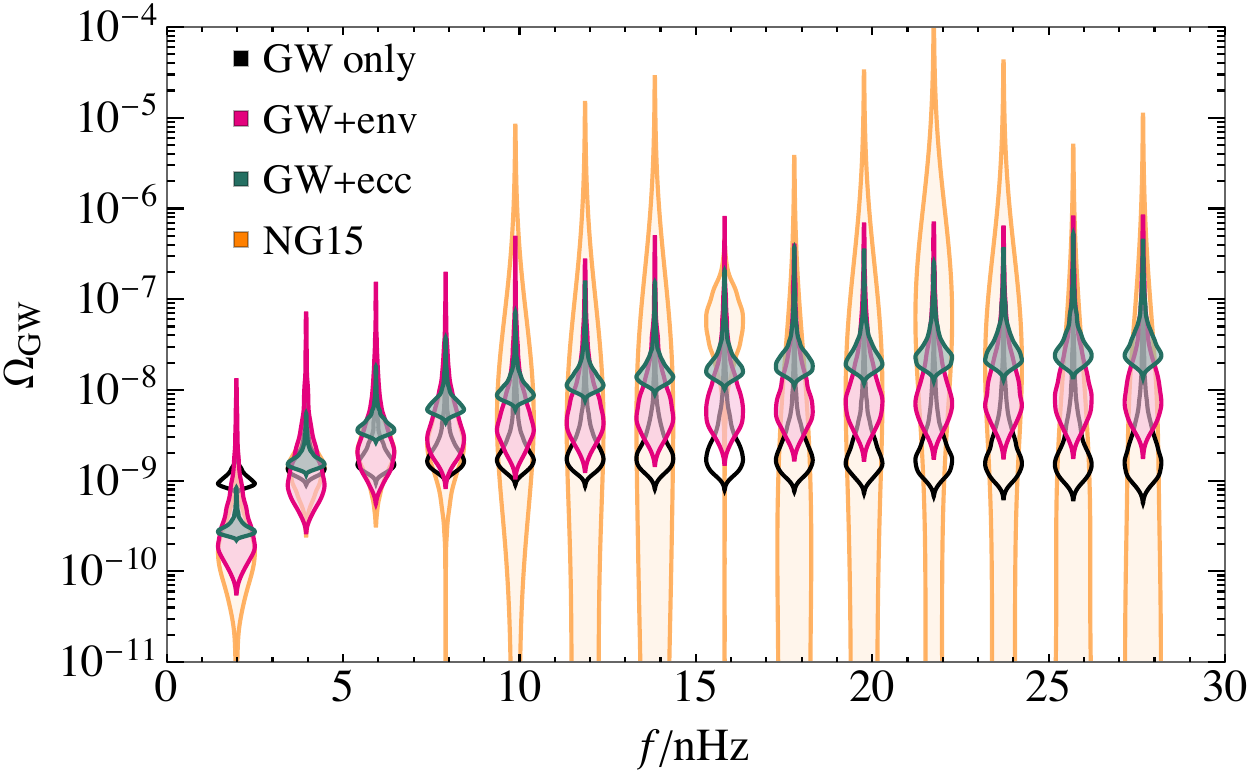}
    \caption{Best fits assuming energy loss by only GWs (black), energy loss by GWs and environmental effects \citep{Ellis:2023dgf} (pink) and energy loss by GWs from eccentric binaries (green) compared to the NG15 data (yellow).}
    \label{fig:bestfit}
\end{figure}

We analysed the model against the NG15 data using the likelihood
\be
    l(\vec{\theta}_f)
    \propto \prod_{i}\int \td \Omega\,  P_{\rm data}(\Omega|f_i) P(\Omega|f_i,\vec{\theta}_f)\, \, ,
\ee
where the product is over the NG15 frequency bins, $P_{\rm data}(\Omega| f_i)$ are the probability distribution functions of the Hellings-Down-correlated free spectrum fit~\citep{NANOGrav:2023gor} (i.e., the NG15 violins shown in Fig.~\ref{fig:bestfit}) and $\vec{\theta}_f$ denotes the model parameters. The full model presented in this paper, including both environmental effects and eccentricity, contains 5 free parameters: $\vec{\theta}_f = \{p_{\rm BH}, \langle e \rangle_{2\,{\rm nHz}}, \alpha, \beta, f_{\rm ref} \}$. Because of the large parameter space and increasingly complex integral for $ P^{(1)}(\Omega|f)$, a full fit of the model to the NG15 data was computationally unfeasible. In the following, we performed fits without environmental effects and with fixed values of $\alpha$, $\beta$ and $f_{\rm ref}$. To compare models we introduced the likelihood ratio
\be
    \ell_{\rm max} = \frac{{\rm max}_{\vec{\theta}_f}l(\vec{\theta}_f |H_1)}{{\rm max}_{\vec{\theta}_f}l(\vec{\theta}_f|H_2)} \,.
\ee
where $H_1$ and $H_2$ label the models.

We performed a scan of the parameter space for the model containing only eccentricity effects, $t_{\rm env}\gg t_{\rm GW}$, with $\vec{\theta}_f = \{p_{\rm BH}, \langle e \rangle_{2\,{\rm nHz}} \}$. The maximal likelihood analysis yielded best-fit values of $p_{\rm BH} = 0.60$ and $\langle e \rangle_{2\,{\rm nHz}} = 0.83$ (corresponding to the power-law index $\gamma_{2 \rm nHz} = 3.9$). The $1\sigma$, $2\sigma$ and $3\sigma$ confidence levels of the two-dimensional likelihood are shown in Fig.~\ref{fig:likelihood} by the coloured regions. The fit prefers values of the SMBH merger efficiency, $p_{\rm BH} > 0.17$ and the likelihood peaks sharply around an initial eccentricity of $\approx 0.83$ due to fitting the low-frequency dip of the NG15 data. At low eccentricities, the contours bend towards the black square, indicating the best fit for circular GW driven binaries. This point is disfavoured by more than $3\sigma$ compared to the fit with eccentric binaries.

For the second scan, we fixed $\alpha = 8/3$, $\beta = 5/8$ and $f_{\rm ref} = 30\,{\rm nHz}$. This choice of $\alpha$ and $\beta$ corresponds to gas infall driven binary evolution~\citep{Begelman:1980vb} and the choice of $f_{\rm ref}$ gives a good fit to the NG15 data assuming circular binaries~\citep{Ellis:2024wdh}. Performing the likelihood analysis in $\vec{\theta}_f = \{p_{\rm BH}, \langle e \rangle_{2\,{\rm nHz}} \}$ yielded the dashed contours seen in Fig.~\ref{fig:likelihood}. The $1\sigma$ region now extends from highly eccentric binaries down to circular binaries. The highest likelihoods are still found at very high eccentricities and near-maximal values of $p_{\rm BH}$. At high eccentricities, the environmental effects are rendered insignificant due to the suppression factor in Eq.~\eqref{eq:ecceffects} and confidence regions become increasingly similar to the scenario without environmental effects.

In Fig.~\ref{fig:bestfit} we show the best-fit spectra of the models with GW-driven circular binaries, circular binaries with environmental effects and GW-driven eccentric binaries. The parameter values are collected in Table~\ref{tab:likelihoods}. The resulting spectrum in the eccentricity model is similar to the spectrum generated by the environmental effects, but the violins are much narrower due to the higher $p_{\rm BH}$ needed. The likelihood ratios reported in Table~\ref{tab:likelihoods} imply that the eccentricity model provides a slightly better fit than the circular binary model with environmental effects.

\begin{table}
\centering
\resizebox{\columnwidth}{!}{%
\small  
\begin{tabular}{cccccc}
    \hline 
    \hline  
    Model & $-2\ln \ell_{\rm max}$ & $p_{\rm BH}$ & $\langle e \rangle_{2\,{\rm nHz}}$ & $f_{\rm ref}/$nHz\\
    \hline
    GW only & 0 (57.5) & $0.06^{+0.01}_{-0.01}$ & 0 & 0\\
    \hline
    GW + env. & -8.8 & $0.37^{+0.09}_{-0.05}$ & $0$ & $30$\\
    GW + ecc. & -12.4 & $0.31_{-0.14}^{+0.40}$ & $0.83_{-0.05}^{+0.04}$ & $0$\\[-1pt]
    & & {\scriptsize $0.60$} & {\scriptsize $0.83$} & {\scriptsize } \\
    GW + ecc. + env. & -10.2 & $0.35_{-0.10}^{+0.53}$ & $0.77_{-0.43}^{+0.07}$ & $30$\\[-1pt]
    & & {\scriptsize $0.90$} & {\scriptsize $0.83$} & {\scriptsize } \\
    \hline
    \vspace{1pt}
\end{tabular} }
\caption{Log-likelihood differences and the posterior means and standard deviations for the considered parametrizations. The best-fit values for the multidimensional fits are given in small print. The log-likelihood differences are shown relative to the GW-only model, for which the absolute log-likelihood value is given in the parenthesis.}
\label{tab:likelihoods}
\end{table}

\section{Conclusions}

We presented a semi-analytic method for computing the stochastic gravitational wave background spectrum generated by a population of inspiraling eccentric supermassive black hole binaries expanding upon the work done in~\citet{Ellis:2023dgf,Ellis:2023owy,Ellis:2024wdh}. Since eccentric binaries spend less time emitting GWs at each frequency, the resulting gravitational wave spectrum of SMBH binaries is generally suppressed depending on the amount of eccentricity present in the SMBH binary population. However, as GW emission tends to circularize orbits, binaries with smaller separations tend to be less eccentric, leading to stronger suppression in the low-frequency tail of the spectrum. We find that this suppression is near-degenerate with the effects of environmental interactions.

We found that the NANOGrav 15-year observations prefer very high initial eccentricities of around $\langle e \rangle_{2\,{\rm nHz}}=0.83$ if the evolution of SMBH binaries in the PTA frequency range is driven purely by gravitational wave emission. When environmental effects are present, the mean eccentricity preferred by data is greatly reduced although non-vanishing eccentricities are still mildly preferred. Our main results are summarized in Fig.~\ref{fig:likelihood}.

Although eccentricity and environmental effects tend to produce very similar spectral modulations, we found they can differ in a few key aspects: (i) The fit of the eccentricity model without environmental effects prefers smaller spectral fluctuations compared to the fit of circular binaries with environmental effects. (ii) As eccentric binaries do not emit monochromatically, they will generate correlations between different frequency bins that can be sizeable for relatively large ($\langle e \rangle_{2\,{\rm nHz}} \gsim 0.5$) initial eccentricity and they cannot be mimicked by environmental effects. These effects provide a potential way to distinguish whether eccentricities or environmental interactions have the dominant effect on the spectral shape of the PTA GW signal.

\vspace{5pt}\noindent{Acknowledgments --}
This work was supported by the Estonian Research Council grants PRG803, PSG869, RVTT3 and RVTT7 and the Center of Excellence program TK202. The work of V.V. was partially supported by the European Union's Horizon Europe research and innovation program under the Marie Sk\l{}odowska-Curie grant agreement No. 101065736.

\bibliography{refs}

\end{document}